\documentclass[10pt]{iopart}

\usepackage{epsfig}

\newcommand{\pizero}{\ensuremath{\pi^0}}
\newcommand{\PT}{\ensuremath{p_T}}
\newcommand{\MT}{\ensuremath{m_T}}

\begin{document}
\title[Hard processes at RHIC]{Overview of Hard processes at RHIC: high-\PT{} light hadron and charm production}

\author{M van Leeuwen}
\address{Lawrence Berkeley National Laboratory, Berkeley, California 94720}
\ead{mvanleeuwen@lbl.gov}


\begin{abstract}
An overview of the experimental results on high-\PT{} light hadron
production and open charm production is presented. Data on particle
production in elementary collisions are compared to next-to-leading
order perturbative QCD calculations. Particle production in
Au+Au collisions is then compared to this baseline.
\end{abstract}
\submitto{\jpg}

\section{Introduction}
The goal of research in high-energy heavy-ion collisions is to study
the properties of strongly interacting matter at extreme energy
density, including the possible phase transition to a
colour-deconfined state: the Quark Gluon Plasma (QGP). The matter
produced in these collisions can be probed using hadrons produced in
partonic processes with a large momentum transfer (`hard
scatterings'). These processes take place early in the collision and
are only sensitive to short distance scales. In the absence of nuclear
effects the hard production yields in nucleus-nucleus collisions are
therefore expected to scale as if the collision were an independent
superposition of nucleon-nucleon collisions. Measurements at SPS have
shown that dilepton production in the Drell-Yan process indeed follows
this expectation \cite{Abreu:1997ji}.

Among the first measurements at RHIC was the measurement of light
hadron production at high transverse momentum \PT{} in Au+Au
collisions, which shows a suppression with respect to the scaled p+p
results. Since these first observations, measurements in d+Au
collisions have confirmed that the observed suppression is a final
state effect. Recently there has also been an increased activity to
verify that high-\PT{} light-hadron production in proton-proton
collisions can be understood in terms of perturbative QCD (pQCD)
calculations, as expected for hard processes. Some of the relevant
results will be reviewed in the next section.

While for light hadron production much of the groundwork
has been done and analyses are clearly moving towards more advanced
observables like identified hadron spectra and correlation
measurements, first results on open charm production are becoming
available. Like high-\PT{} light hadrons, open charm is expected to be
dominantly produced in hard processes and can therefore serve as a
calibrated probe of the medium. Due to their large mass, however,
charm quarks and hadrons are expected to be affected differently by
the medium than high-\PT{} light hadrons \cite{molnar_sqm04}.

In the second part of this paper an overview will be presented of the
existing results on open charm production at RHIC. The present results
are based on run-2 and run-3 data. First results from the large
statistics Au+Au data sample from run-4 are to be expected soon. These
will greatly improve the precision and \PT-coverage of the open charm
measurements in Au+Au collisions at RHIC.

\section{High-\PT{} light hadron production}
High-\PT{} hadron production at is the most readily
accessible observable for hard processes at RHIC. At sufficiently
high \PT, all hadrons are expected to be produced in jet
fragmentation. The non-perturbative dynamics of jet-fragmentation can
be characterised by a universal fragmentation function $D(z)$ which is
parametrised using data from $e^+e^-$ collisions at different energies
\cite{Kniehl:2000fe,Kretzer:2000yf}. With these fragmentation
functions and the parton densities from deep inelastic
scattering experiments, the expected cross sections for high-\PT{} hadron
production can be calculated in perturbative QCD (pQCD).

\subsection{Neutral pions and charged hadrons in p+p collisions}

\begin{figure}
\begin{minipage}{0.49\textwidth}
  \epsfig{file=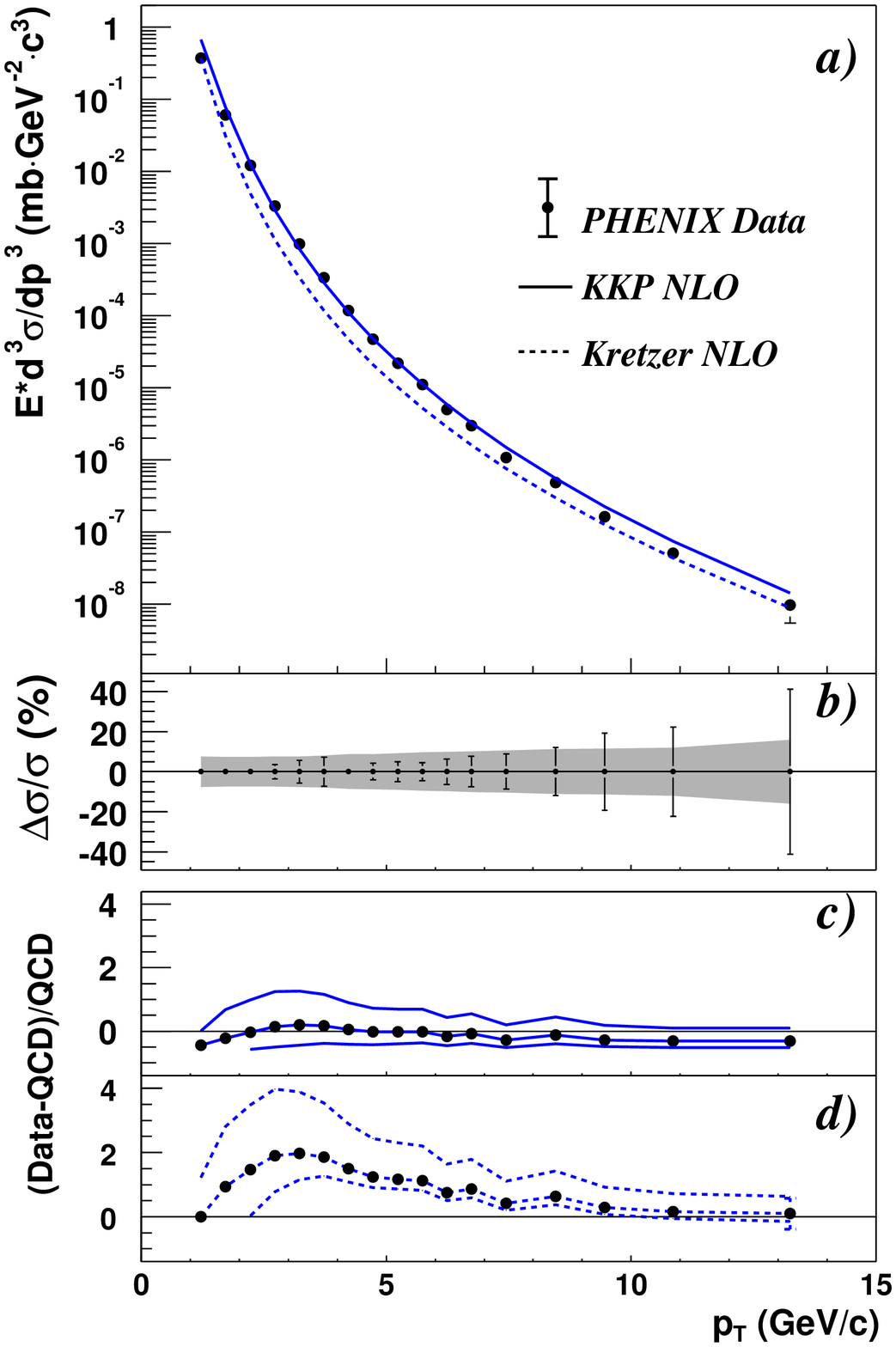,width=\textwidth}
\end{minipage}\hfill
\begin{minipage}{0.49\textwidth}
\epsfig{file=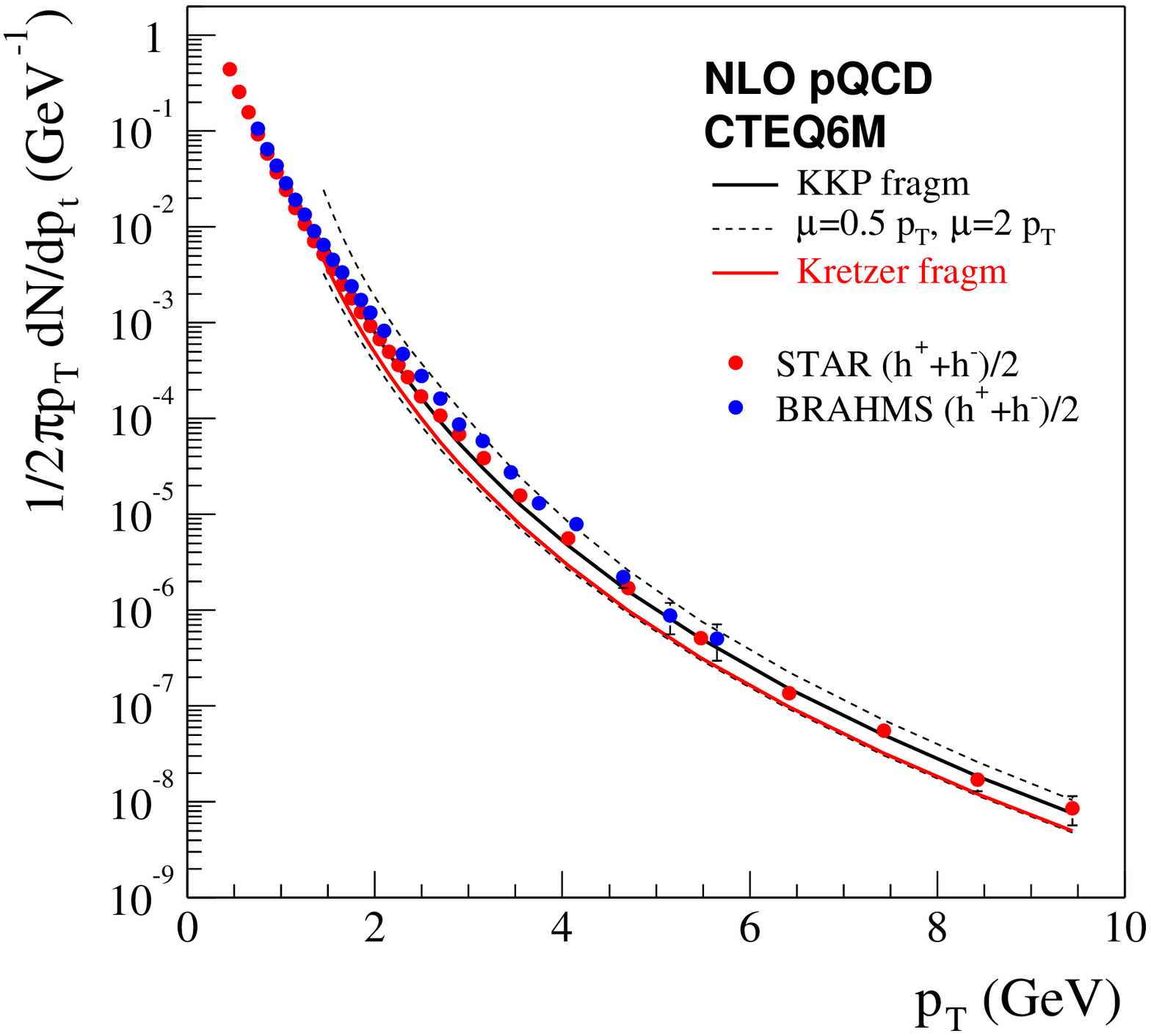,width=\textwidth}
\epsfig{file=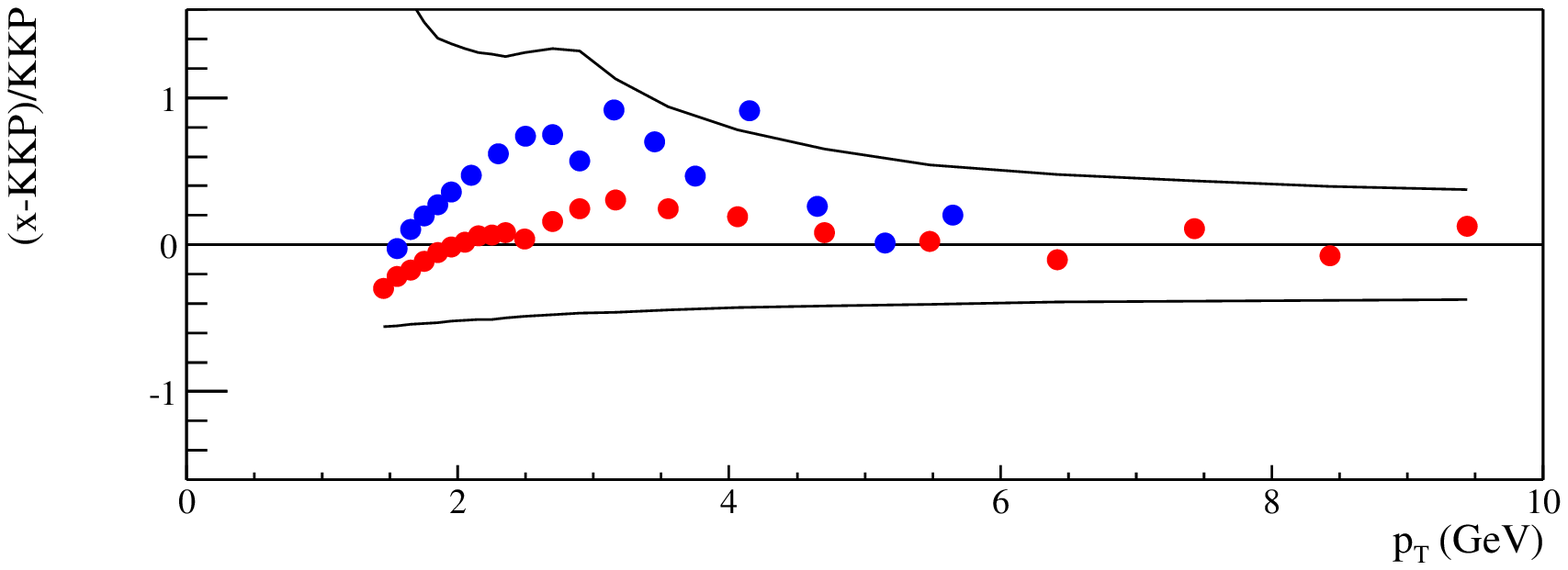,width=\textwidth}
\end{minipage}
  \caption{\label{lh_nlo}Measured \PT-distributions of \pizero{}
  (left) and charged hadrons (right) in p+p collision compared to NLO
  pQCD calculations (lines). The different lines indicate results
  using different parametrisations of the fragmentation functions (see
  text) and choice of the factorisation and renormalisation
  scales. The lower panels show the difference between data and
  calculations, divided by the calculated curves.}
\end{figure}

\Fref{lh_nlo} shows \PT-spectra of \pizero{} measured by PHENIX
\cite{Adler:2003pb} (left panel) and charged hadrons from STAR
\cite{Adams:2003kv} and BRAHMS \cite{Arsene:2004ux} (right panel) in
p+p collisions at $\sqrt{s}=200$~GeV. The data are compared to a
next-to-leading order (NLO) pQCD calculation \cite{Jager:2002xm}. The
uncertainty in the theoretical calculation is estimated by varying the
renormalisation and factorisation scales $\mu_R$ and $\mu_F$ to half
and twice the nominal value of $\mu_R=\mu_F=\PT$. These scale
variations change the calculated cross-sections by about 20\% for
$\PT>5$~GeV. To illustrate the uncertainty in the fragmentation
functions, the calculation was performed with two different sets of
fragmentation functions, from Kniehl, Kramer and Potter (KKP)
\cite{Kniehl:2000fe} and from Kretzer \cite{Kretzer:2000yf}. Both sets were
independently determined from similar selections of $e^+e^-$ data
using slightly different assumptions about relations between the
fragmentation functions for different partons. This turns out to be the
dominant source of uncertainty for the \pizero{} spectrum: variations
of the order of 50\% are seen, mainly due to uncertainties in the
gluon fragmentation function. Note, however, that these are partly
normalisation uncertainties and do not change the shape of the spectra
very much.  The measurements have an overall normalisation uncertainty
of about 10\%, which is not indicated in the figures. The systematic
offset between the STAR and BRAHMS results in \fref{lh_nlo} can
probably be attributed to this normalisation uncertainty.

Both for neutral pions and charged hadrons, data and theory agree over
more than 5 orders of magnitude, which gives confidence that hadron
production at high \PT{} ($>3$~GeV) is indeed governed by hard point-like
processes.

\subsection{Strange hadron production in p+p collisions}
The above comparisons can be extended to the strange hadrons $K^0_S$ and
$\Lambda$. While the kaon fragmentation function is relatively
well-constrained by the data from $e^+e^-$ collisions, data on
$\Lambda$ production are scarce
\cite{deFlorian:1997zj}.

\begin{figure}
  \centering
  \epsfig{file=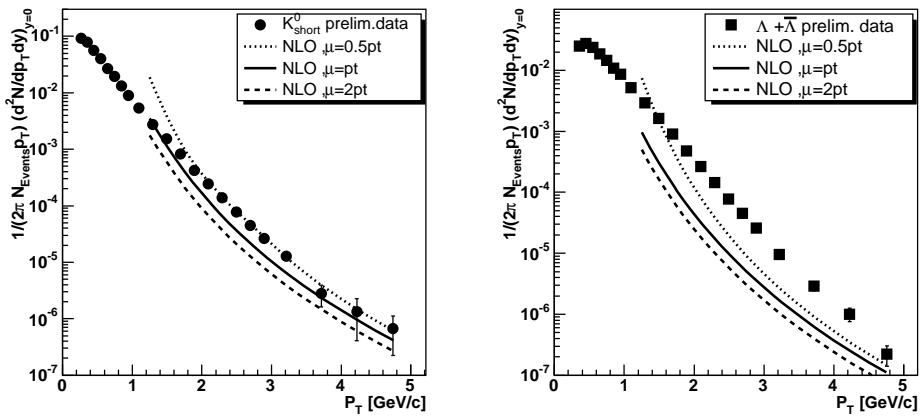,width=\textwidth}
  \caption{\label{k0_lam}Comparison of $K_S^0$ and $\Lambda$ spectra in
  p+p collisions as measured by STAR to NLO pQCD calculations
  \cite{heinz_sqm04}. The different lines indicate results for
  several settings of the factorisation and normalisation scales.}
\end{figure}

A first comparison of $K^0_S$ and $\Lambda$ spectra in
$\sqrt{s}=200$~GeV p+p collisions to NLO calculations as
presented at this conference is shown in \fref{k0_lam}
\cite{heinz_sqm04}.  The agreement between the measured $K^0_S$ spectrum
and the expectation from NLO pQCD is reasonable, although the shape of
the calculated spectrum is slightly more concave than the measured
one. This difference is mainly apparent at relatively low \PT{} (1-2
GeV), where soft production processes may still contribute
significantly.

For the $\Lambda$ on the other hand, the agreement between the data and
the NLO calculations is far from satisfactory. This might be
indicative of the breakdown of the massless formalism and the
factorization ansatz for particles with mass that is significant
compared to $\PT$~\cite{Kretzer:2004ie}. Before drawing this
conclusion, however, the uncertainties in the $\Lambda$ fragmentation
functions should be better quantified.

\subsection{Suppression in Au+Au collisions}
\begin{figure}
  \epsfig{file=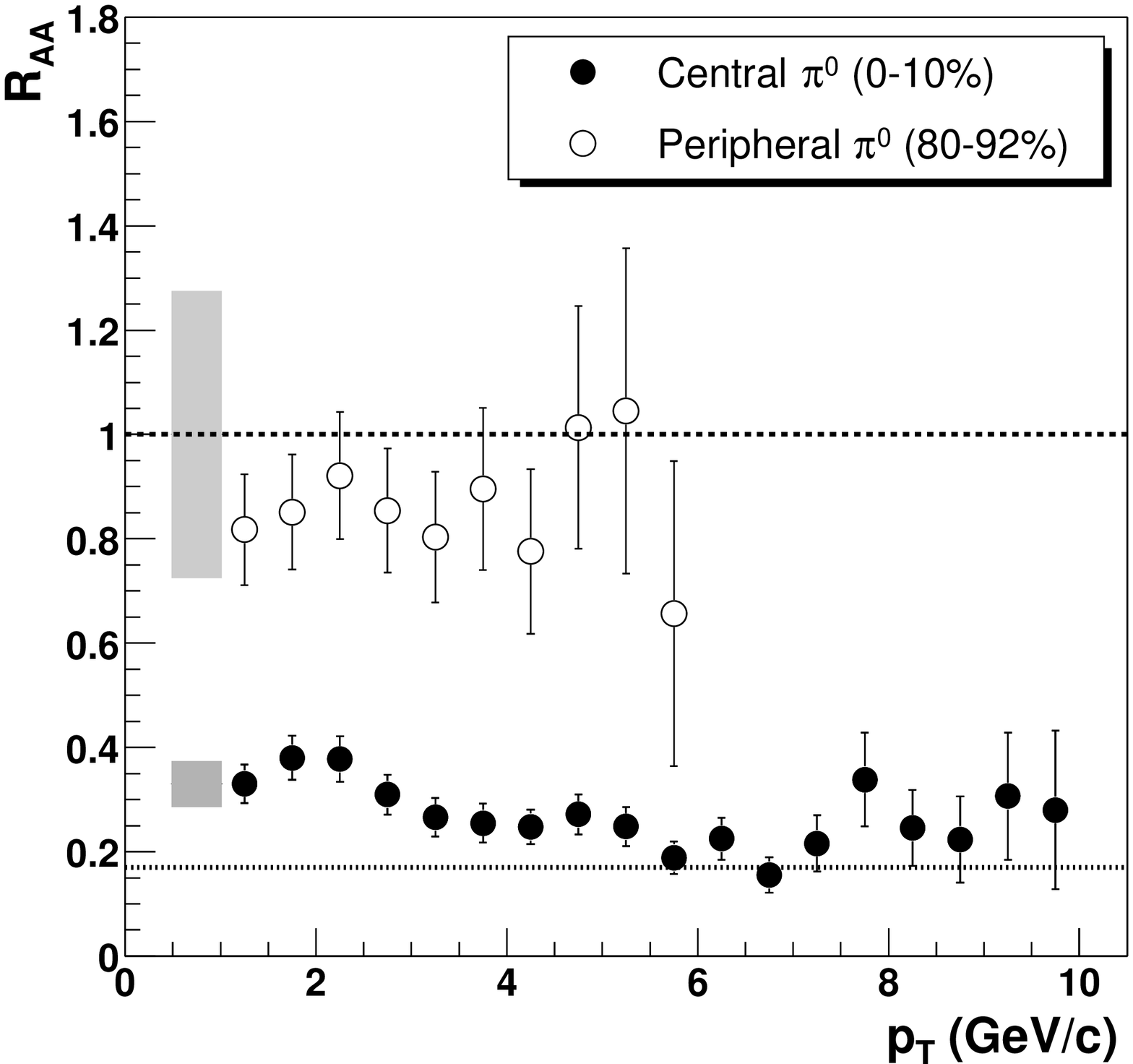,width=0.42\textwidth}\hfill
  \epsfig{file=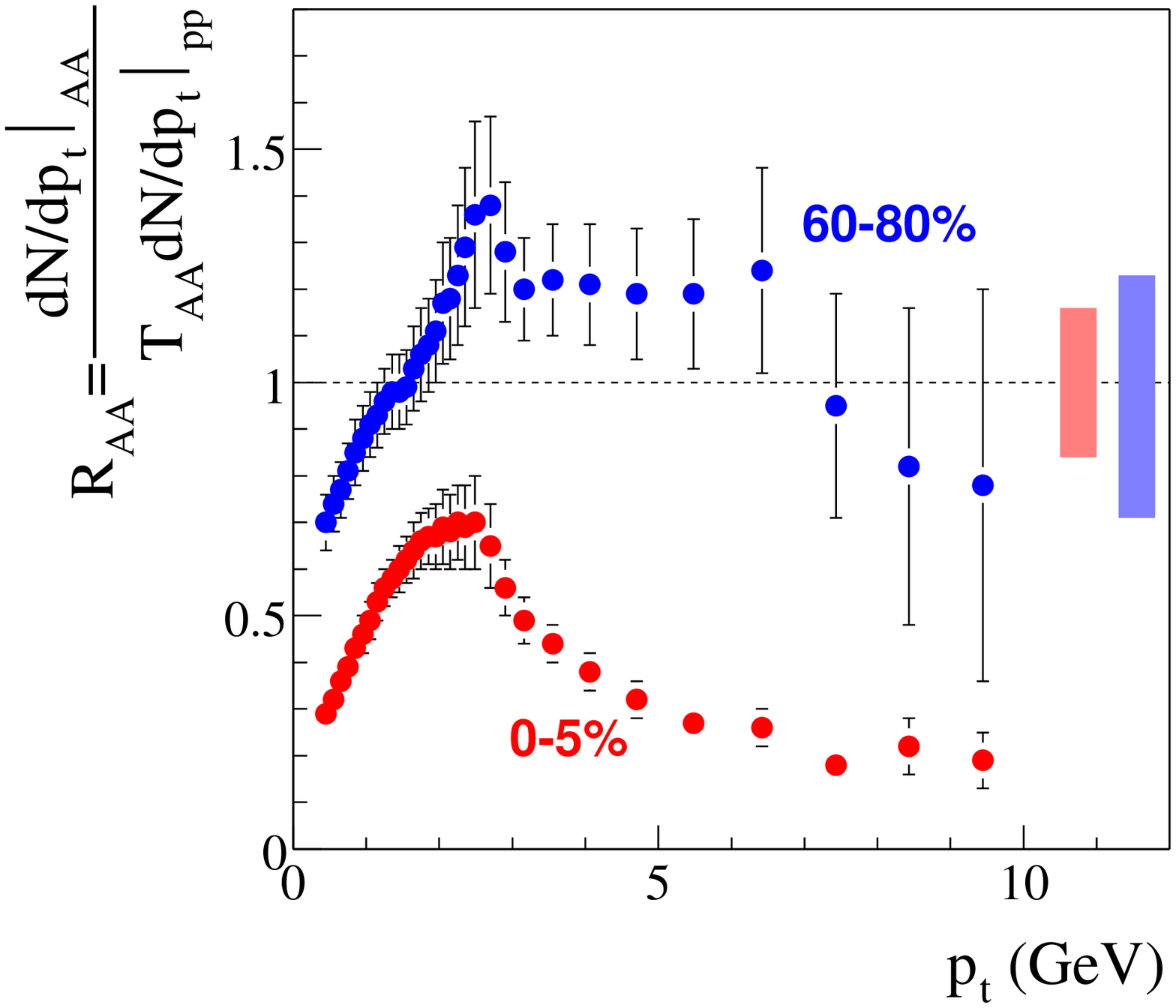,width=0.50\textwidth}
  \caption{\label{raa}Nuclear modification factor $R_{AA}$ for \pizero{}
  and charged hadron production in central and peripheral Au+Au
  collisions, as measured by PHENIX \cite{Adler:2003qi} and STAR
  \cite{Adams:2003kv}.}
\end{figure}
To compare measured the particle spectra in Au+Au collisions to the expected
$N_{coll}$ scaling from p+p, the nuclear modification factor 
\begin{equation}
R_{AA}=  \frac{\left.dN/d\PT\right|_{Au+Au}}{N_{coll}
  \left.dN/d\PT\right|_{p+p}}
\end{equation}
is generally used. In \fref{raa} these ratios are shown for peripheral
and central Au+Au collisions at $\sqrt{s}=200$~GeV, both for \pizero{}
from PHENIX (left panel) \cite{Adler:2003qi} and charged hadrons from
STAR (right panel) \cite{Adams:2003kv}.  The suppression ratio
$R_{AA}$ in peripheral collisions is close to unity, while for central
collisions a suppression of up to a factor 5 is observed. This
suppression was one of the first indications of a strong final state
modification of particle production in Au+Au collisions that is now
generally ascribed to energy loss of the fragmenting parton in the hot
and dense medium.

\section{Open charm production in d+Au and Au+Au collisions}
While light hadron production is only expected to be calculable in perturbative
QCD at higher \PT, the charm quark mass ($m_c\approx 1.35$~GeV) is
large enough to expect pQCD calculations to be valid for all
\PT. Final state effects on charm quarks and hadrons are expected to
be smaller than for the light hadrons due to the large charm
mass~\cite{molnar_sqm04}.

\subsection{Charmed meson spectra in d+Au collisions}

\begin{figure}
\epsfig{file=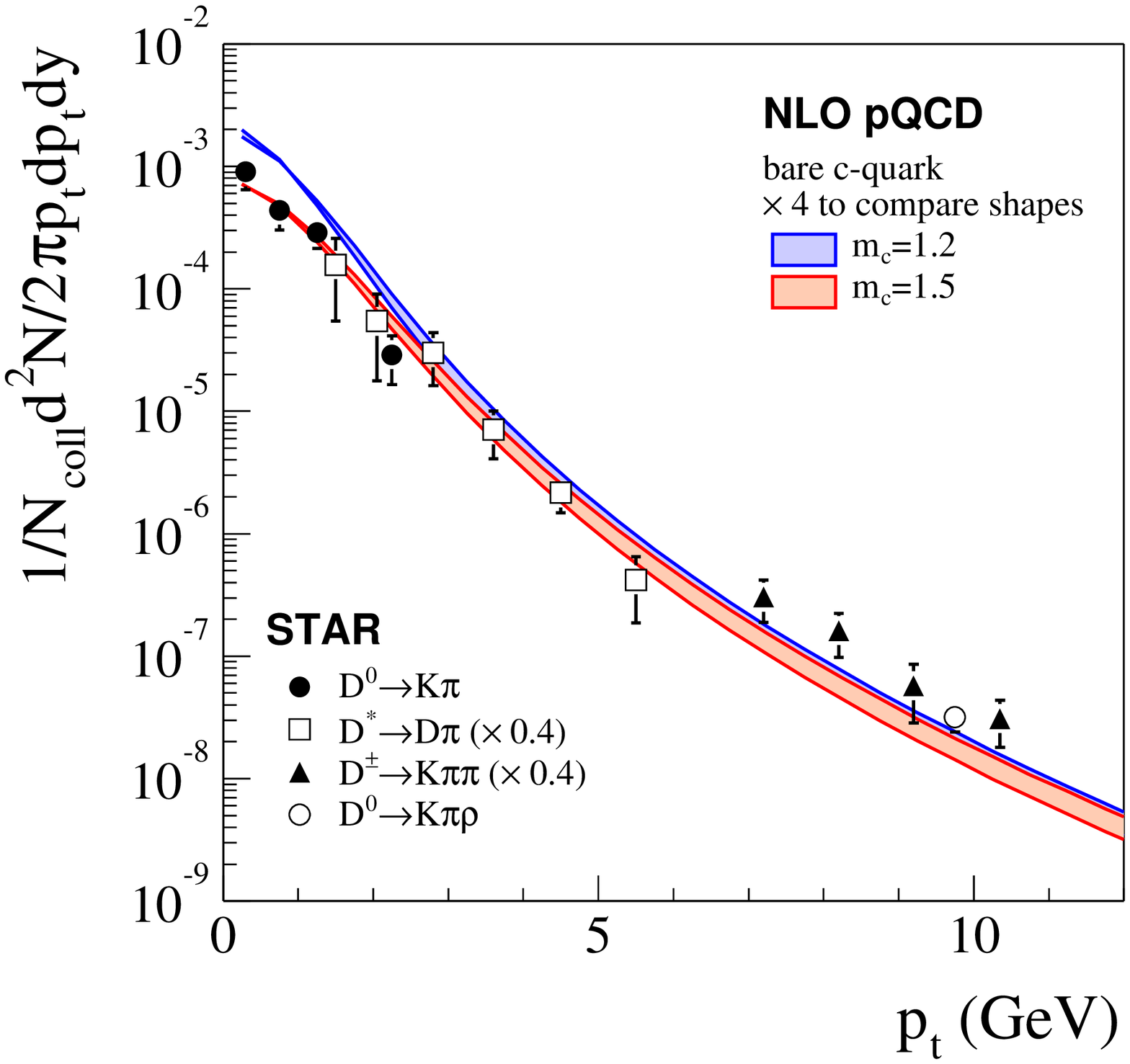,width=0.45\textwidth}\hfill
\epsfig{file=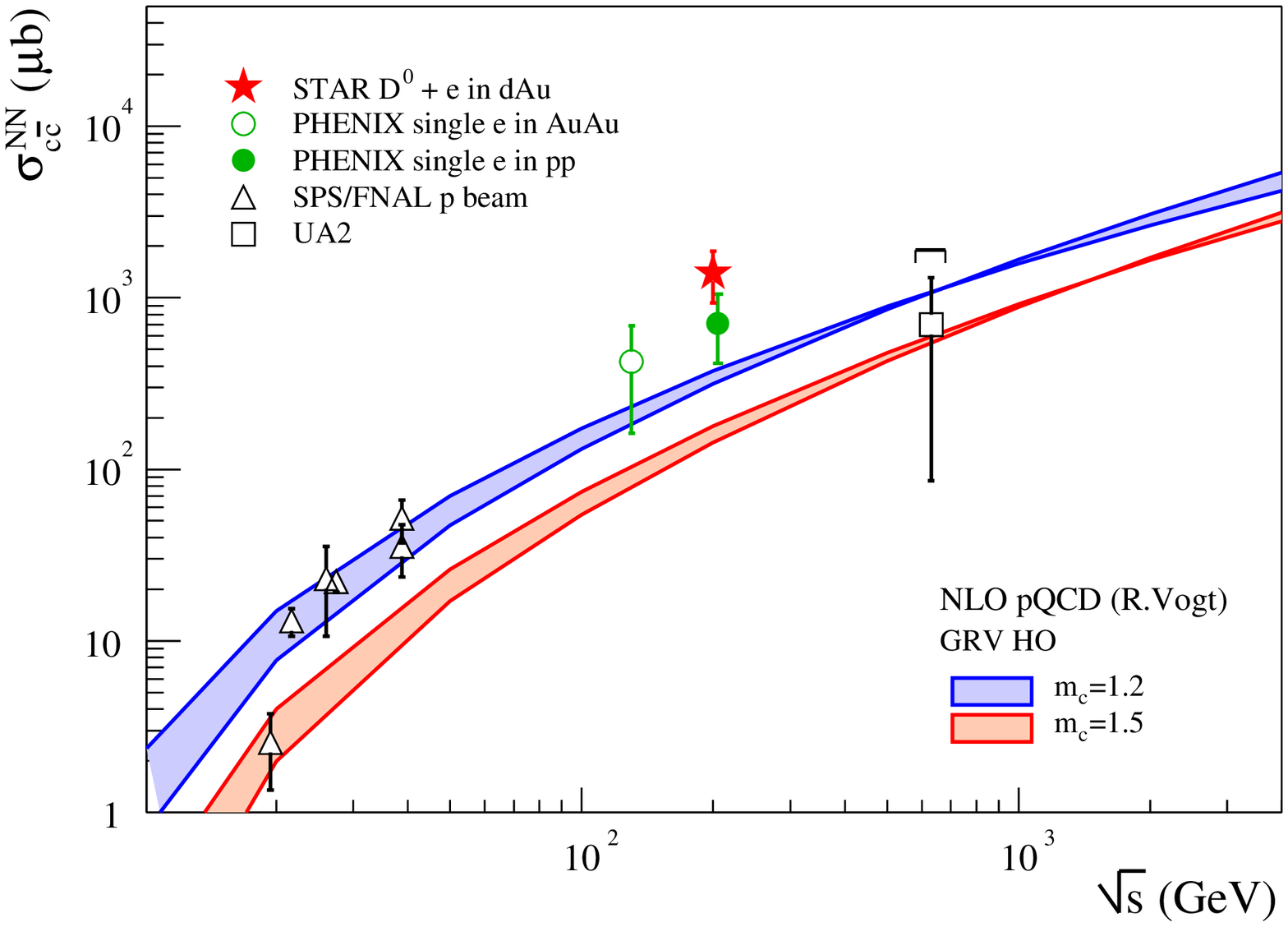,width=0.45\textwidth}
\caption{(left) \label{charm_dau} Collected \PT{} spectra for $D^0$,
  $D^{\pm}$ and $D^{*\pm}$ production in d+Au collisions at
  $\sqrt{s}=200$~GeV. The spectra have been divided by
  $N_{coll}\approx7$ to be comparable to calculations for p+p. The
  $D^{\pm}$ and $D^{*\pm}$ spectra were scaled to match the $D^0$
  data. (right) Energy dependence of the total cross-section for charm
  production \cite{Adams:2004fc,Kelly:2004qw}. The different symbols
  indicate different experiments and collision systems. All data were
  scaled to effective p+p cross sections. When available, the
  statistical and systematic error were added in quadrature. The bands
  in both panels show NLO pQCD calculations \cite{Vogt:2001nh} with
  two different assumptions for the charm mass and two choices of the
  factorisation and normalisation scales (see test).
  variations. }
\end{figure}
None of the RHIC experiments currently has a vertex detector with
sufficient resolution to reconstruct secondary vertices of charm
decays. Even without secondary vertex reconstruction, STAR has been
able to statistically reconstruct decays to charged hadrons in d+Au
collisions at $\sqrt{s}=200$~GeV using an invariant mass method
\cite{Tai:2004bf}. Different charmed mesons ($D^0$, $D^{\pm}$, and
$D^{*\pm}$) can be reconstructed in different \PT{} ranges, thus
providing a charm measurement up to $\PT{}=11$~GeV, as shown in
the left hand panel of \fref{charm_dau}. The $D^\pm$ (triangles) and
$D^{*\pm}$ (squares) spectra have been scaled to match the $D^0$
spectra (circles). Note that the $D^0$ has been measured at low \PT{}
and at high \PT{} using different decay modes and thus provides a
normalisation for the whole \PT{} range.

Also shown in \fref{charm_dau} (left) are NLO pQCD calculations of
{\it charm quark spectra} \cite{Vogt:2001nh}. To illustrate the
sensitivity of those calculations to the charm quark mass $m_c$ and
the choice of renormalisation scale, curves are drawn for
$m_c=1.2$~GeV and $m_c=1.5$~GeV and with two choices for the
renormalisation and factorisation scales: $\mu_R=\mu_F=\MT$ and
$\mu_R=\mu_F=2\MT$. Note that the shape of the spectra at low \PT{} is
most sensitive to both the charm quark mass and the choice of scales.

For a detailed comparison of the data to theory, the calculated
charm quark spectrum should be convoluted with the charm fragmentation
function. This would lead to a softening of the spectrum and a
reduction of the yield, both of which may in principle depend on the
meson species. Given the limited \PT{} range of the spectra for the
separate species and the relatively large uncertainties in their
fragmentation functions, we have chosen to compare the shape of the
combined meson spectrum directly to the charm quark spectra. For this
purpose, the calculated spectra were scaled up by a factor of 4 to
approximately match the data. The shapes of the calculated charm quark
spectra and the measured meson spectra are surprisingly similar,
leaving little room for softening due to fragmentation.

All in all, it is far from clear that the present data can be matched
with a NLO pQCD calculation. Before drawing any conclusions about
charm production in elementary collisions at RHIC, though, we should
wait until the present data are finalised. Although it is expected
that $N_{coll}$ scaling is valid for charm production in d+Au, it
would be good to confirm this by similar measurements in p+p. There
are also some open questions for theory. For example, a matched
next-to-leading logarithm calculation is needed to describe beauty
production at the Tevatron \cite{Cacciari:2003uh}. In addition, a new
way of extracting fragmentation functions, by fitting the Mellin
moments instead of a direct $z$-space fit is found to lead to an
effectively harder fragmentation function
\cite{Cacciari:2002pa}. Similar considerations may also affect
calculations of charm production at RHIC.

\subsection{Total charm cross section}
The right-hand panel of \fref{charm_dau} shows the measured energy
dependence of the total charm quark cross section
\cite{Adams:2004fc,Kelly:2004qw}, compared to NLO pQCD calculations
\cite{Vogt:2001nh}. Estimating the total charm quark cross section
from experimental data involves substantial extrapolations to the
unmeasured regions of momentum space and corrections to include
unmeasured charmed hadron species. These corrections lead to sizeable
systematic uncertainties on the data, as can be seen from the
figure. The uncertainties on the NLO pQCD calculations due to higher
order corrections and the choice of the charm mass are also
significant. It is therefore preferable to directly compare data and
calculations in the measured regions (see also
\cite{Frixione:2004md}).

Note also that it seems that the charm cross section per
nucleon-nucleon collision as measured by STAR in d+Au collisions from
a combination of the electron measurements and the invariant mass
method is somewhat higher than expected from the trend observed by
other experiments and the pQCD calculations. The deviations are within
the present uncertainties on the measurements.

\subsection{Centrality dependence of charm production in Au+Au collisions}
\begin{figure}
  \epsfig{file=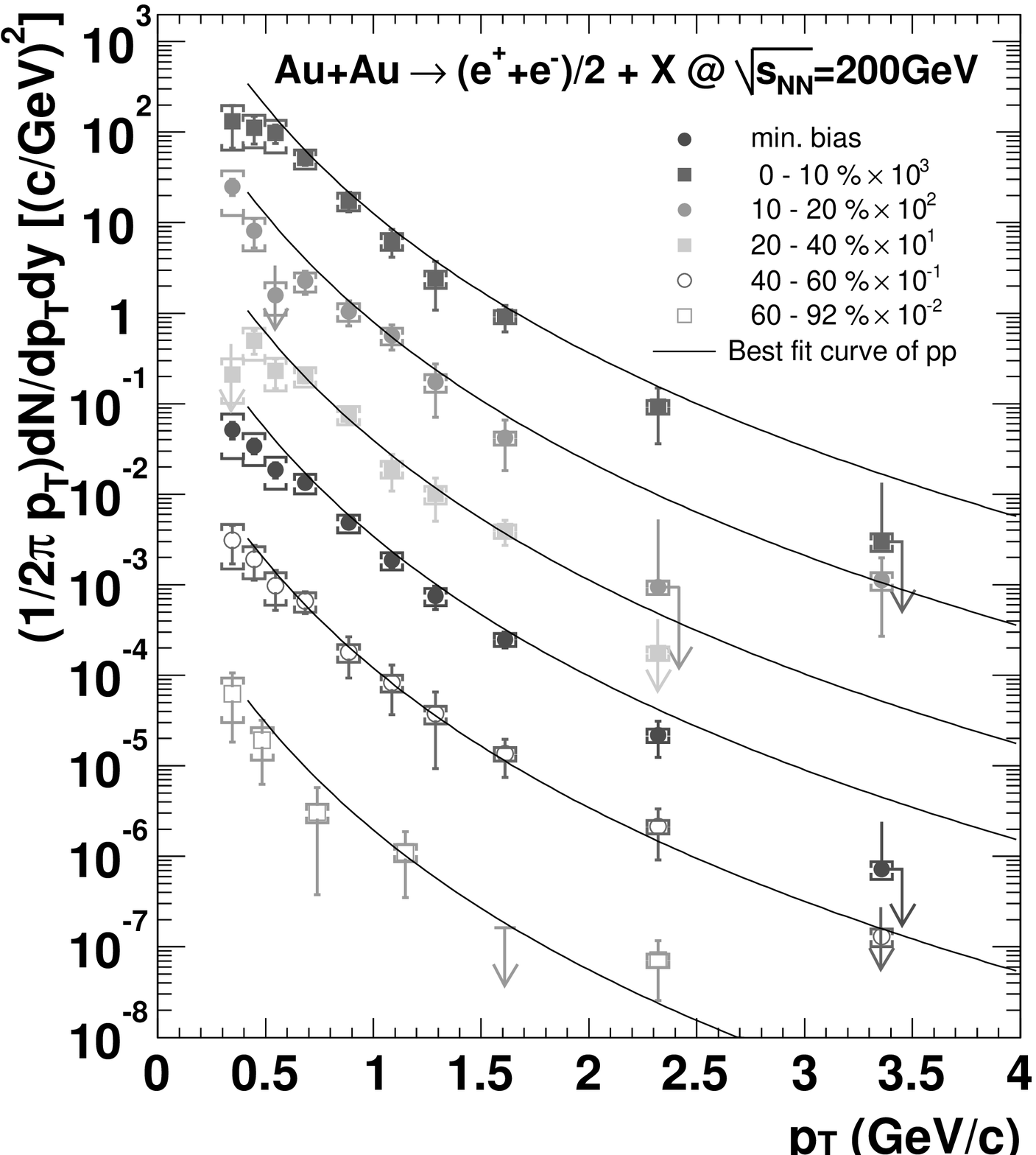,width=0.45\textwidth}
  \hfill\epsfig{file=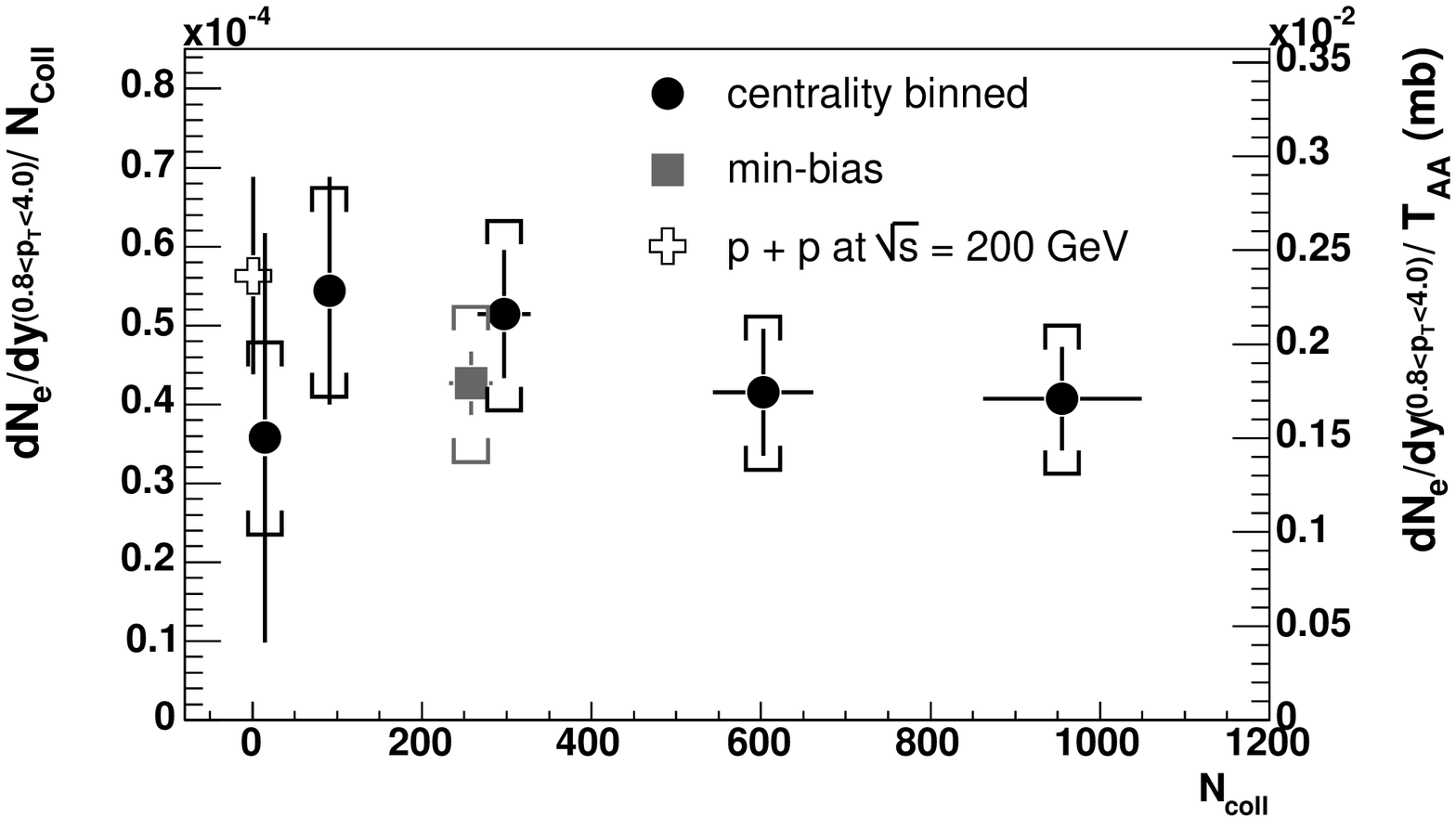,width=0.53\textwidth}
  \caption{\label{electron_auau} (left) Non-photonic electron
  spectra from Au+Au collisions, compared to the expectation from
  p+p. (right) Nuclear modification factors $R_{AA}$ for
  non-photonic electrons with $\PT>0.8$~GeV as measured by
  PHENIX~\cite{Adler:2004ta}.}
\end{figure}
A first indication of the centrality dependence of charm production in
Au+Au collisions can be taken from electron spectra. After subtraction
of the contributions from light hadrons (mainly through photon conversions,
but also from Dalitz decays of \pizero, $\eta$, $\eta'$, $\rho$,
$\omega$ and $\phi$) the electrons from heavy flavour decays remain
(`non-photonic' electrons). In the left
hand panel of \fref{electron_auau}, the electron spectra from heavy
flavour decays as measured by PHENIX  in centrality-selected Au+Au collisions \cite{Adler:2004ta} are shown. The
lines show reference spectra obtained from a fit to the measured
spectrum in p+p, scaled by the number of collisions. At each
centrality, the spectra agree with the expected $N_{coll}$ scaling
from p+p, albeit within large errors.

In the right-hand panel of \fref{electron_auau}, the yields of
non-photonic electrons with $0.8<\PT<4.0$~GeV per nucleon-nucleon
collision are shown as a function of centrality. There is no
indication of a suppression as seen for light hadrons
(see \fref{raa}). One should keep in mind, however, that electrons with
$\PT>0.8$~GeV have contributions from semi-leptonic charm decays at
all \PT. The presented results are therefore not very sensitive to a
possible suppression of charm production at moderate or high \PT{}
($>2$~GeV).

\subsection{Charm flow}
\begin{figure}
  \begin{centering}
  \epsfig{file=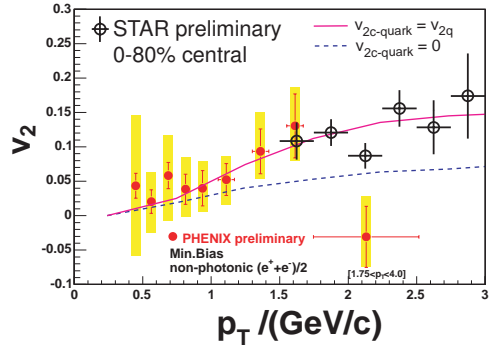,width=0.5\textwidth}
  \caption{\label{charm_flow} Elliptic flow $v_2$ of non-photonic
  electrons in Au+Au collisions as measured by STAR and PHENIX~\cite{laue_sqm04}.}
  \end{centering}
\end{figure}

A measurement of the elliptic flow $v_2$ of charmed mesons is an
independent way of assessing the sensitivity of charmed mesons to
final state interactions. Here again, we have to rely on measurements
of decay electrons for the time being. In \fref{charm_flow} the
elliptic flow of non-photonic electrons is shown
\cite{laue_sqm04}. Both STAR and PHENIX observe non-zero electron
flow, which is a strong indication that charmed mesons flow. This is
an intriguing possibility, because it would show decisively that charm
production is sensitive to the dense hadronic or even partonic
environment in the collision. At the moment the statistical and
systematic errors are still large, precluding a precise quantitative
extraction of flow values. The situation is expected to dramatically
improve with the larger Au+Au data samples which were recorded
this year.

\section{Summary and outlook}
A comparison of neutral pion and charged hadron \PT-spectra measured
in p+p collisions at $\sqrt{s}=200$~GeV at RHIC to NLO pQCD
calculations shows that high-\PT{} light hadron production is well
described by perturbative QCD, albeit within relatively large
uncertainties, mainly from the fragmentation functions. This gives
confidence that high-\PT{} particle production is governed by hard,
point-like processes, for which the cross section in Au+Au collisions
is expected to scale with the number of nucleon-nucleon collisions. A 
suppression of light hadrons by approximately a factor of 5 compared to
the expected scaling is observed in central Au+Au collisions, due to
final state interactions of the fragmenting quarks and/or the produced
hadrons.

For strange hadrons, $K_S^0$ and $\Lambda$, the agreement between data
and NLO pQCD calculations is not as good, or even unsatisfactory (for
$\Lambda$). In those cases, however, the data do not extend to very
high \PT{} and the fragmentation functions are not as well known as
for the light hadrons. This needs more investigation before
conclusions can be drawn about the applicability of pQCD.

The same is true for open charm production in d+Au collisions, where
shape of the measured $D$ meson spectra is similar to the calculated
charm quark spectra, leaving little room for softening due to
fragmentation.

First results on electron production from PHENIX indicate that there
is no or very little suppression of charm production in Au+Au
collisions. Measurements of electron flow, on the other hand,
indicate significant flow of the charmed mesons, which can only be
due to significant final state interactions.

In the near future, a measurement of the \PT-dependence of nuclear
modification factors for non-photonic electrons or maybe even open
charm can be expected from the large statistics Au+Au data samples
collected in run-4 at RHIC. This, together with a more accurate
measurement of charm flow, will map out the interactions of
charm quarks and hadrons with the medium and, through comparison with the
light hadron results, may eventually shed more light on the nature
of these interactions.

\end{document}